\documentstyle[prl,aps,epsfig,floats]{revtex}

\draft
\begin{document}

\newcommand{\gsim}{\lower.7ex\hbox{$\;\stackrel{\textstyle>}{\sim}\;$}}
\newcommand{\lsim}{\lower.7ex\hbox{$\;\stackrel{\textstyle<}{\sim}\;$}}
\newcommand{\mev}{\hbox{\rm\,MeV}}
\newcommand{\gev}{\hbox{\rm\,GeV}}
\newcommand{\tev}{\hbox{\rm\,TeV}}
\newcommand{\xpb}{\hbox{\rm\, pb}}
\newcommand{\xfb}{\hbox{\rm\, fb}}

\def\NPB#1#2#3{Nucl. Phys. B {\bf #1}, #3 (19#2)}
\def\PLB#1#2#3{Phys. Lett. B {\bf #1}, #3 (19#2)}
\def\PLBold#1#2#3{Phys. Lett. B {\bf#1}, #3 (19#2)}
\def\PRD#1#2#3{Phys. Rev. D {\bf #1}, #3 (19#2)}
\def\PRL#1#2#3{Phys. Rev. Lett. {\bf#1}, #3 (19#2)}
\def\PRT#1#2#3{Phys. Rep. {\bf#1}, #3 (19#2)}
\def\ARAA#1#2#3{Ann. Rev. Astron. Astrophys. {\bf#1}, #3 (19#2)}
\def\ARNP#1#2#3{Ann. Rev. Nucl. Part. Sci. {\bf#1}, #3 (19#2)}
\def\MODA#1#2#3{Mod. Phys. Lett. A {\bf #1}, #3 (19#2)}
\def\ZPC#1#2#3{Zeit. f\"ur Physik C {\bf #1}, #3 (19#2)}
\def\APJ#1#2#3{Ap. J. {\bf#1}, #3 (19#2)}
\def\MPL#1#2#3{Mod. Phys. Lett. A {\bf #1}, #3 (19#2)}

\twocolumn[\hsize\textwidth\columnwidth\hsize\csname
@twocolumnfalse\endcsname

\title{Massive Doublet Leptons}
\author{Scott Thomas${}^a$ and James D. Wells${}^b$}
\address{${}^{(a)}$Physics Department, Stanford University, 
                             Stanford, CA 94305 \\
${}^{(b)}$Stanford Linear Accelerator Center, Stanford University,
   Stanford, CA 94309}
\date{April 17, 1998}
\maketitle

\begin{abstract}
Massive vector-like electroweak doublets
are generic in many extensions of the standard model.
Even though one member of the doublet is necessarily
electrically charged these particles are not easily
detected in collider experiments.
The neutral and charged states within the doublet
are split by electroweak symmetry breaking.
In the absence of mixing with other states, the
radiatively generated splitting is in the range
$200 - 350\mev$ for $m \gsim {1 \over 2} m_Z$.
The charged state decays to the neutral one
with an ${\cal O}$(cm) decay length,
predominantly by emission of a soft charged pion.
The best possibility to detect these massive charged particles
is to trigger on hard initial state radiation and search for
two associated soft charged pions with displaced vertices.
The mass reach for this process at LEPII
is limited by luminosity rather than kinematics.

\end{abstract}
\pacs{PACS numbers: 14.60.Hi, 13.35.-r, 13.85.Qk
\hfill SLAC-PUB-7799, SU-ITP 98-22 \hfill hep-ph/9804359}

]

The best probe for physics beyond the standard model
is direct production of new states at high energy colliders.
Many theories of electroweak symmetry breaking
require additional states which are charged under
electroweak gauge interactions.
In many cases the states form chiral representations
of $SU(2)_L \times U(1)_Y$, and necessarily gain
mass only from electroweak symmetry breaking.
However, it is possible for states with electroweak scale
mass to transform under vector representations,
even though there may be no symmetry apparent in the
low energy
theory which protects them from gaining a large mass.
For example, the masses of vector-like fermions can
be protected by global chiral symmetries which are
spontaneously broken at the electroweak scale.
Likewise,
in supersymmetric theories, matter supermultiplets which transform
under a vector representation, and are massless
at the high scale, remain massless to all orders due to the
non-renormalization theorem.
The fermionic components of
such supermultiplets can gain mass from a field which is
a singlet under $SU(2)_L \times U(1)_Y$, but nonetheless gains
an expectation value in association with electroweak symmetry
breaking.

Massive vector representations can naturally carry a
conserved or approximately conserved quantum number.
This can forbid or highly suppress mixing with standard model fermions,
and render the lightest state of the representation
effectively stable on the scale of an accelerator experiment.

In this paper we discuss the phenomenology of a
massive stable vector fermion doublet of $SU(2)_L$. 
In grand unified theories
this representation arises in ${\bf 5} \oplus \bar{\bf 5} \in  SU(5)$
or equivalently if a standard model generation is embedded
in ${\bf 27} \in E_6$.
Vector representations of this type may also be required
in theories of low scale supersymmetry breaking in which
$U(1)_{PQ}$ and $U(1)_{R-PQ}$ Higgs sector symmetries
are spontaneously broken at the electroweak scale \cite{GMSB}.
Furthermore, the fermionic partners of the up- and down-type
Higgs bosons in supersymmetry form a vector-like $SU(2)_L$ doublet.
These Higgsinos become mass eigenstates
in the limit  $|m_{\lambda}^2 - \mu^2 | \gg m_Z^2$,
where $m_{\lambda}$ are the gaugino masses.
The analysis given below becomes applicable in this limit
if the Higgsino is the lightest supersymmetric particle.

Surprisingly, even though one member of the doublet is
necessarily charged, these states turn out to be
very difficult to detect experimentally.  
It is usually assumed that if kinematically
accessible, a heavy charged particle is easily detected
at a high energy 
collider.
This is generally true if (i) the heavy charged particle
is non-relativistic and lives long enough to pass through
the entire detector, depositing a greater than minimum ionizing
track, or (ii) it decays promptly to visible final states with
energetic charged leptons and/or jets.
The electrically charged state of the
vector-like $SU(2)_L$ doublet discussed here
satisfies neither (i) nor (ii).
The decay length to the neutral state,
although macroscopic, is too short to allow
direct triggering on the primary charged tracks.
In addition, the visible decay products are too soft to allow direct
triggering.
However, as discussed below, triggering on associated initial state
radiation allows a search for decays over a macroscopic distance
to the very soft charged particles in the final state.

The representation considered here is a pair of $SU(2)_L$
doublet Weyl fermions with $U(1)_Y$ hypercharge $Y= \pm  1$,
where $Q = T_3 + {1 \over 2} Y$ \cite{early}.
This may be represented as a single Dirac fermion
\begin{equation}
L = \left( \begin{array}{c}
L^0 \\
L^-
\end{array} \right)_{Y= -1} ~.
\end{equation}
Other hypercharge assignments are not unifiable in a conventional
manner, and have both members of the doublet charged.
This representation is referred to as a doublet lepton
since the left-handed component has the same gauge quantum numbers
as a left-handed standard model lepton.
This Dirac state can gain an $SU(2)_L \times U(1)_Y$ invariant
mass, ${\cal L} \supset - m\bar L L= - m(\bar L^+ L^-+\bar L^0 L^0)$.
In the absence of mixing with standard model leptons,
the lowest order operator which can split $L^-$ and $L^0$
in the presence of $SU(2)_L \times U(1)_Y$ breaking is
$\bar{L} T^a L ~ H^{\dagger} T^a H$, where $H$ is
the Higgs boson operator.
For fermionic doublets this is a non-renormalizable operator.
In a renormalizable theory it receives finite calculable corrections.
The mass splitting
$\delta m \equiv m_{L^{\pm}} - m_{L^0}$
is therefore calculable within the low energy
theory \cite{susynote}.

\begin{figure}
\centerline{\epsfig{figure=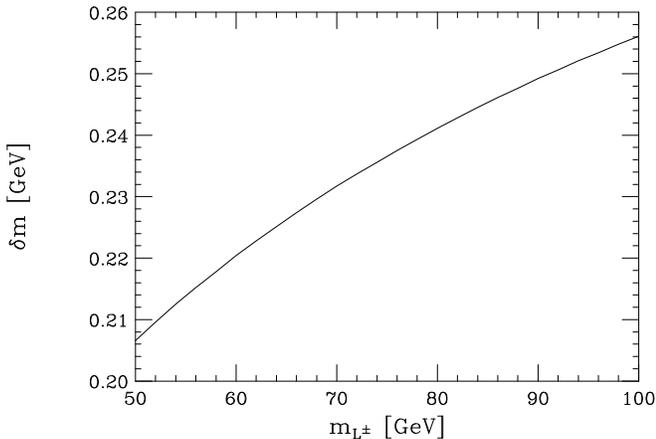,width=3.4in,angle=0}}
\caption{
{\small
Doublet mass splitting
$\delta m \equiv m_{L^{\pm}} - m_{L^0}$
as a function of  $m_L$.}}
\label{dm}
\end{figure}

At lowest order the mass splitting comes from one-loop corrections
with virtual photon and $Z$ boson exchange to both the masses and
wave functions.
Virtual $W^\pm$ bosons do not contribute 
since the couplings to $L^-$ and $L^0$ are identical.
The one-loop mass splitting
for on-shell states is
\begin{equation}
\delta m = {\alpha \over 2} m_Z f(m_L^2 / m_Z^2)
\label{spliteq}
\end{equation}
where $f(r)$ is the loop function
\begin{equation}
f(r) = \frac{\sqrt{r}}{\pi} ~ \int_0^1 dx~
\left( 2 - x \right) {\rm ln} \left[ 1 + \frac{x}{r (1-x)^2} \right] .
\end{equation} 
For $r \ll 1$, $f(r) \rightarrow 0$ and for $r \gg 1$,
$f(r) \rightarrow 1$.
The radiatively generated mass splitting
is plotted in Fig.~\ref{dm} for $m_L$ in the range 50--100 GeV.
The asymptotic value of the splitting for $m_{L}^2 \gg m_Z^2$
is $\delta m = {1 \over 2} \alpha  m_Z \simeq 355$ MeV.
In this limit the mass renormalization is twice as large
in magnitude and opposite in sign as compared with wave function
renormalization.

The important features of the radiatively induced mass splitting can
be understood in an effective field theory analysis.
In the low energy theory below $m_Z$ the $L^-$ mass receives
a linearly divergent contribution from the virtual photon loop.
This divergence is cut off in the full theory
by momenta above ${\cal O}(m_Z)$ for which electroweak symmetry
is effectively restored.
The splitting is therefore proportional to the electromagnetic fine
structure constant times the $Z$ boson mass.
The linear divergence in momentum space corresponds in
real space to the Coulomb self energy of $L^-$.
In the heavy field limit of $m_L^2 \gg m_Z^2$, the mass splitting
(\ref{spliteq}) is precisely the difference between the Coulomb
self energies of $L^-$ and $L^0$
due to the photon and $Z$ boson classical electric
fields \cite{scalarsplit}.
In this interpretation it is clear that
$L^-$ is heavier than $L^0$, and that the splitting vanishes without
electroweak symmetry breaking.

The form of the splitting can also be understood in the effective theory above
$m_Z$.
In this description the coupling of the gauge eigenstates $W^3$ and $B$,
of the $SU(2)_L$ and $U(1)_Y$ gauge groups respectively, to $L^-$ and
$L^0$ are identical.
Gauge invariance then implies that
only diagrams which mix $W^3$ and $B$ through
an even number of Higgs insertions can
contribute to the splitting.
All these effective operators
receive infrared divergent contributions which are cut off
by momenta of ${\cal O}(m_Z)$.



In a supersymmetric theory there are additional contributions
to the mass splitting (\ref{spliteq}).
At lowest order these come from one-loop diagrams with internal
neutralinos and the scalar partner of the vector doublet.
With an $SU(2)_L \times U(1)_Y$ invariant soft mass for the scalar
partner of the form ${\cal L} = - m_{\tilde{L}}^2
\tilde{L}^{\dagger} \tilde{L}$, these contributions appear only as corrections
to the vector doublet wave function.
Electroweak symmetry breaking enters the supersymmetric loops at lowest
order in two ways.
The first is through gaugino-Higgsino mixing in the neutralino mass
matrix.
Since the lowest order operator in the effective theory
above $m_Z$  which splits $L^0$ and $L^-$
requires at least two Higgs insertions, this contribution arises only
at second order in gaugino-Higgsino mixing.
In the mostly gaugino or Higgsino region of parameter space this contribution
is then suppressed compared with (\ref{spliteq}) by
${\cal O}(m_L m_Z / (\mu^2 - m_{\lambda}^2))$.
The second way electroweak symmetry enters is through the scalar partner
$SU(2)_L$ $D$-term.
This splits the scalar $\tilde{L}^-$ and $\tilde{L}^0$ masses by
${\cal O}(m_Z^2 / m_L)$.
At one loop this modifies the vector doublet splitting by an amount
which is suppressed compared with (\ref{spliteq}) by
${\cal O}(m_L m_Z / m_{\tilde{L}}^2)$.
The loop momenta for both types of supersymmetric contributions are
${\cal O}({\rm max}(m_{\tilde{L}}, m_{\lambda}))$.
Because of these inherent suppressions, over much of the parameter
space possible supersymmetric contributions are small compared
with the dominant standard model contribution (\ref{spliteq})
to the doublet mass splitting.
Small corrections are however sensitive to the superpartner spectrum.

\begin{figure}
\centerline{\epsfig{figure=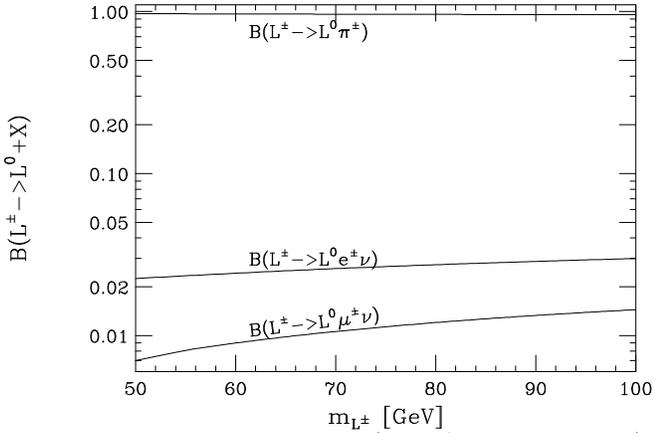,width=3.4in,angle=0}}
\caption{
{\small
Branching ratios for $L^{\pm} \to L^0 X$ where
$X=\pi^{\pm}$, $e^{\pm} \nu$, or $\mu^{\pm} \nu$.}}
\label{bratio}
\end{figure}

The neutral state of the doublet, $L^0$, is rendered effectively
stable by discrete or continuous global chiral symmetries.
The state $L^-$ can however decay to $L^0$ via charged current
interactions.
For the mass range of interest here the most important decay
modes are
$L^{\pm} \to L^0 \pi^{\pm}$, $L^0 e^{\pm} \nu$, and
$L^0\mu^{\pm} \nu$.
The partial widths for these modes are
\begin{eqnarray}
\Gamma(L^{\pm} \to L^0 \pi^{\pm}) & = &
        \frac{G_F^2}{\pi} \cos^2 \theta_c  f_\pi^2
        \delta m^3 \sqrt{1-b_\pi^2}
                \label{piondecay} \\
\Gamma(L^{\pm} \to L^0 l^{\pm} \nu) & = &
        \frac{G_F^2}{15\pi^3}\delta m^5
         \sqrt{1-b^2_l} P(b_l)
                 \label{leptondecay}
\end{eqnarray}
where,
\begin{equation}
P(b_l)= 1-\frac{9}{2}b_l^2-4b_l^4+
   \frac{15b_l^4}{2\sqrt{1-b_l^2}}
  {\rm tanh}^{-1}\sqrt{1-b_l^2},
\end{equation}
$f_\pi\simeq 130\mev$, $\theta_c$ is the Cabibbo angle,
$b_\pi =m_{\pi}/\delta m$, and
$b_l =m_l/\delta m$.
The branching ratios for each of these modes are plotted
in Fig.~\ref{bratio} as a function of $m_{L}$.
\noindent
The exclusive  mode $L^{\pm} \to L^0\pi^-$
of course dominates since it is two body and
since $f_{\pi} \sim \delta m$.

Observable signatures of a massive vector doublet
are very limited.
Virtual contributions to the oblique electroweak
parameters 
are insignificant since the doublet does not gain a mass
from electroweak symmetry breaking.
The $S$ parameter is proportional to corrections to
mixing between $W^3$ and $B$ gauge eigenstates.
This requires at least two Higgs insertions and arises only
at two loops.
The $T$ parameter is proportional to isospin violation,
which likewise arises only at two loops.
Direct decay of the $Z$ boson to massive doublets
is however important if kinematically open.
The contribution to the $Z$ boson total width
is equivalent to $2(1 + (1- 2 \sin^2 \theta_W)^2) \simeq 2.6$
massive Majorana neutrino species.
This would unacceptably modify the $Z$ width
unless $m_L \gsim {1 \over 2} m_Z$.



\begin{figure}
\centerline{\epsfig{figure=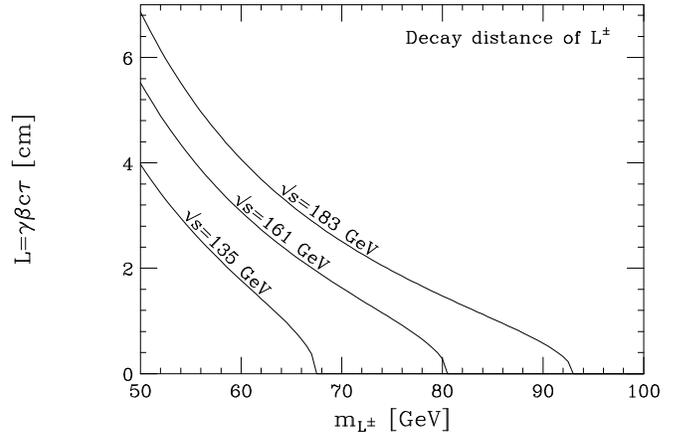,width=3.4in,angle=0}}
\caption{
{\small
Decay distance of $L^{\pm}$ in the lab frame
at an $e^+e^-$ collider
as a function of $m_{L}$.
The decays are boosted for $\sqrt{s}=135$,
$161$, and $183$ GeV.}}
\label{gbctau}
\end{figure}

Direct detection of doublets which are too heavy to affect the
$Z$ boson total width is very challenging
even though they are produced
copiously if kinematically accessible;
at an $e^+e^-$ collider,
$\sigma(L^+L^-) \sim \sigma(L^0 \bar{L}^0) \sim
\sigma(\mu^+\mu^-)
\sqrt{1 - 4 m_L^2 / s}$ \cite{djouadi}.
The neutral $L^0$ and $\bar{L}^0$ interact weakly
like a massive neutrino and exit
the detector without depositing visible energy.
The principle reason for the difficulty in observing
$L^{\pm}$ is that the decays (\ref{piondecay})
and (\ref{leptondecay}) give an $L^{\pm}$  decay length
of ${\cal O}({\rm cm})$.
The lab frame $L^{\pm}$ decay distance
for different center of mass energies relevant at LEPII are shown in
Fig.~\ref{gbctau}
\noindent
The typical decay length is unfortunately too short to utilize
a topological trigger which identifies essentially back-to-back
charged tracks largely independent of total energy deposition
in the detector.
Such a trigger requires that at least one of the tracks
traverse the inner tracking region which typically extends
to ${\cal O}(30~{\rm cm})$.
Triggering on the very soft charged decay products is equally
difficult.
For the $L^{\pm} \to L^0 \pi^{\pm}$ decay mode, the $\pi^{\pm}$
radius of curvature in the detector magnetic field is
${\cal O}({\rm m / Tesla})$.
Separating such tracks at the trigger level
from soft charged tracks arising from beam-beam
interactions is problematic.

\begin{figure}
\centerline{\epsfig{figure=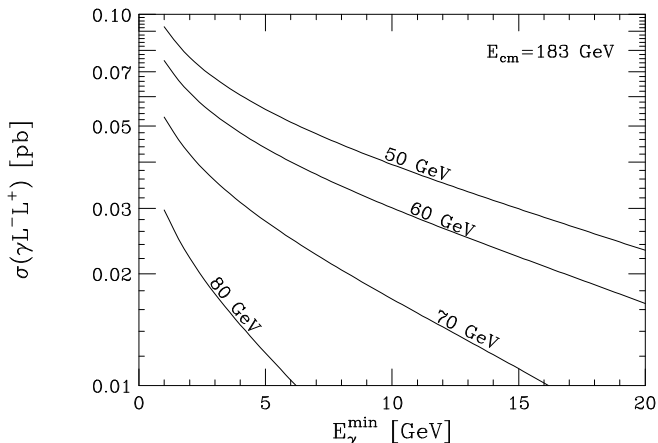,width=3.4in,angle=0}}
\caption{
{\small
Total cross-section
$\sigma(e^+e^-\to L^+L^- \gamma)$ at $\sqrt{s} = 183\gev$
with $|\cos \theta_\gamma|<0.7$ and $E_\gamma>E^{\rm min}_\gamma$
as a function of $E^{\rm min}_\gamma$
for various values of $m_{L}$.}}
\label{rad183}
\end{figure}

One method to search for production of invisible or nearly invisible
particles is to trigger on an associated hard radiated photon.
This has been suggested for counting neutrino species \cite{neutrino},
and as a means to search for neutral supersymmetric particles, including
photinos \cite{photino,phosneutrino},
neutralinos \cite{vlsp,neutralino}
sneutrinos \cite{phosneutrino,vlsp,sneutrino}, and nearly degenerate
Higgsinos or Winos \cite{higgsino}.
In the approximation that the associated photon
arises solely from initial state radiation, a photon
radiator function \cite{radiator}
can be convoluted with the radiation free cross section to obtain
the differential cross as a
function of $c_\gamma\equiv \cos \theta_\gamma$ and
$x_\gamma =E_\gamma /E_{\rm beam}$:
\begin{equation}
\frac{d\sigma(L^+L^- \gamma)}{dx_\gamma dc_\gamma} =
\sigma({L^+L^-})((1-x_\gamma)s)R(x_\gamma,c_\gamma;s)
\end{equation}
where,
\begin{equation}
R(x_\gamma,c_\gamma;s)=\frac{\alpha}{\pi}\frac{1}{x_\gamma}
\left[ \frac{1+(1-x_\gamma)^2}{1+4m^2/s-c_\gamma^2} -
\frac{x^2_\gamma}{2}\right] .
\end{equation}
The LEP experiments can trigger on
central photons with $|\cos \theta_{\gamma}| \lsim 0.7 $ and
energies greater than 5--10 GeV \cite{opal}.
The cross section $\sigma(e^+e^-\to L^+L^- \gamma)$
at $\sqrt{s}=183\gev$ with this
photon coverage is plotted
in Fig.~\ref{rad183} for several values of $m_L$
as function of the minimum photon energy
for tagging, $E_{\gamma}^{\rm min}$.

The largest backgrounds for single hard photon production
are $e^+ e^- \to \nu \bar{\nu} \gamma$ and $e^+e^-\to Z\gamma$
with $Z\to \nu\bar\nu$, and to a lesser extent
$e^+ e^- \to l^+ l^- \gamma$ with both $l^+$ and $l^-$
forward and undetected.
However, at the analysis level these can be separated from the signal
$e^+ e^- \to L^+ L^- \gamma$ by requiring identification of two
soft $\pi^{\pm}$ and/or $l^{\mp}$ arising from $L^{\pm}$ decays.
Additional processes with
very soft $\pi^{\pm}$ and/or $l^{\mp}$ in association with a photon
may provide a small background.
These can be separated from the signal by requiring that the
soft tracks have some non-vanishing impact parameter with
the beam axis.
Pairs of soft charged tracks with displaced vertices
in association with a hard photon provide a striking signature
for massive vector doublets.

With an integrated luminosity of $240 \xpb^{-1}$
between the four LEP experiments at $\sqrt{s}=183\gev$,
we estimate that the analysis described above could be sensitive
to a doublet mass up to roughly 70 GeV.
This assumes 5 signal events with $E_{\gamma}^{\rm min} = 8$ GeV and
$|\cos \theta_{\gamma}| < 0.7$.
A full Monte Carlo simulation of
the signal and background with complete detector performance
folded in would probably reduce this reach slightly.

Unlike many other signatures at $e^+e^-$ machines,
the experimental reach for vector-like lepton doublets
is not limited merely by the center of mass energy, but rather by the
luminosity of the accelerator.  Future LEPII runs with higher energy
and much higher luminosity will greatly increase the search capability
for these particles.
Extension of searches to future lepton colliders such as the NLC
is also possible.
Searches for $L^+ L^-$ production in association with a
photon or $Z$ boson are also in principle possible at
hadron colliders.
The larger background of soft charged tracks from beam-beam interactions
however makes identification of displaced charged tracks from $L^{\pm}$
decay more challenging.

Finally, it is interesting to note that
independent precision measurements of $m_L$ from the total
cross section, and of $\delta m$ from the $L^{\pm}$ decay length
distribution and decay pion energy spectrum
would be sensitive to deviations from the standard model
one-loop mass splitting (\ref{spliteq}).
This would provide an indirect probe for additional states
beyond the photon and $Z$ boson which can couple the heavy doublet
to electroweak symmetry breaking through virtual processes,
such as in supersymmetric theories.

\bigskip
\noindent
{\it Acknowledgements: }
We thank S.~Dimopoulos and A.~Litke for many enlightening
discussions.  We also
thank N.~Arkani-Hamed, G.~Landsberg and A.~Sopczak for helpful conversations.
The work of S.T. was supported by Stanford University through
a Fredrick E. Terman fellowship and of J.D.W. by the Department
of Energy under contract DE-AC03-76SF00515.


\begin{thebibliography}{99}
\bibitem{susynote}
For a scalar doublet, the operator
$L^{\dagger} T^a L ~ H^{\dagger} T^a H$ is renormalizable,
and the splitting between a scalar $L^-$ and $L^0$ is
an a priori free parameter of the theory.
In supersymmetric theories these operators are
$SU(2)_L$ $D$-terms.
\bibitem{scalarsplit} Since spin decouples in the heavy field
limit the radiative correction to the mass splitting for
very massive scalar doublets is identical to that of fermion doublets.
\end{thebibliography}
\end{document}